\documentclass[aps,prl,twocolumn]{revtex4}

\usepackage{graphicx}
\usepackage{subfigure}
\usepackage{verbatim}
\usepackage{color}
\usepackage{amsmath}

\begin{document}

\title{Exploiting classical nucleation theory for reverse self-assembly}
\author{William L. Miller}
\author{Angelo Cacciuto}
\email{ac2822@columbia.edu}
\affiliation{Department of Chemistry, Columbia University, \\New York, New York 10027}
\date{\today}
\begin{abstract}
In this paper we introduce a new method to design interparticle interactions to target arbitrary crystal structures via the process of self-assembly.
We show that it is possible   to exploit the curvature of the crystal nucleation free-energy barrier to sample and select optimal 
interparticle interactions for self-assembly into a desired structure. 
We apply this method to find interactions to target two simple crystal structures: a crystal with simple cubic symmetry
and a two-dimensional plane with square symmetry embedded in a three-dimensional space. 
Finally, we discuss the potential and limits of our method and 
propose a general model by which a functionally infinite number of different interaction geometries may be constructed 
and to which our reverse self-assembly method could in principle be applied. 
\end{abstract}
 
\maketitle
\section*{INTRODUCTION}

Not only is understanding, controlling and predicting the phenomenological behavior of particle self-assembly one of the great mathematical challenges for the 21st century, but its applications in materials science and engineering hold promise for the development of materials with novel electronic, mechanical, and optical properties.
Although most of the work performed in this field is historically rooted in the self-assembly of small molecules, the last decade has witnessed extraordinary advances in particle synthesis at the meso-scale~\cite{DeVries,Schnablegger,Hong,Weller,Hobbie}, making possible the production of  building blocks  with complex chemical and geometrical properties with an unprecedented degree of precision.
Unfortunately, a coherent theoretical framework around the problem of self-assembly is still missing, and numerical simulations have taken the lead in exploring the wealth of new phenomenological behavior arising from the collective behavior of non-isotropic components.

Most  numerical studies on self-assembly of nanoparticles performed so far have adopted the patchy sphere model~\cite{Kern}.
In this model, the isotropicity of a particle is broken by placing on its surface regions (patches) with different physical properties; for example, hydrophobic chemical groups or single-stranded DNA chains.
Theoretically, these regions are incorporated into the inter-particle potential by a simple angular dependence which favors or disfavors the alignment of such patches. 
Although self-assembly of several simple  structures has been achieved with the patchy models (references \cite{glotzer4,Liu,cacciuto}  are just a few examples of the large body of work published on the subject; for a recent review see references~\cite{sciortino1,sciortino2}), a general modeling approach to the problem is missing.
Shape and position of the interaction sites is either guessed using physical arguments, or inspired by known molecules or protein structures aggregating into a similar target crystal.
There are two notable exceptions: the inverse optimization technique proposed by Torquato~\cite{torquato}, which is specific to nondirectional interactions, and so-called ``bottom-up building block assembly,'' devised by Jankowski and Glotzer~\cite{glotzer}, which requires the construction of the 
most relevant terms of the partition function of the system, starting from individual particles.
The development of an efficient numerical procedure to design interactions between nanoparticles that targets specific crystal structures via the process of self-assembly would therefore be a result of great importance. 

Although the generic features of particle aggregation can be described, at least phenomenologically, in terms of simple thermodynamic arguments~\cite{Israelachvili, Zhang, Leckband, Nagarajan, glotzer2}, the details of the self-assembly process are far from being understood, even in the simple case of aggregation of isotropic particles into macroscopic three-dimensional crystals.
In fact, a full theoretical description of this problem must incorporate critical kinetic effects which are not captured by classical thermodynamics, and which have dramatic macroscopic consequences~\cite{chandler, cacciuto,
whitelam}.
 
It is now understood that for self-assembly to take place, a very delicate balance between entropic and energetic contributions, coupled to a precise geometric character of the components, must be satisfied. 
In general, self-assembly of nanocomponents is not to be expected unless a careful design of the building blocks has been performed beforehand.~\cite{chandler, cacciuto} 

\begin{figure}
	\includegraphics[width=0.44\textwidth]{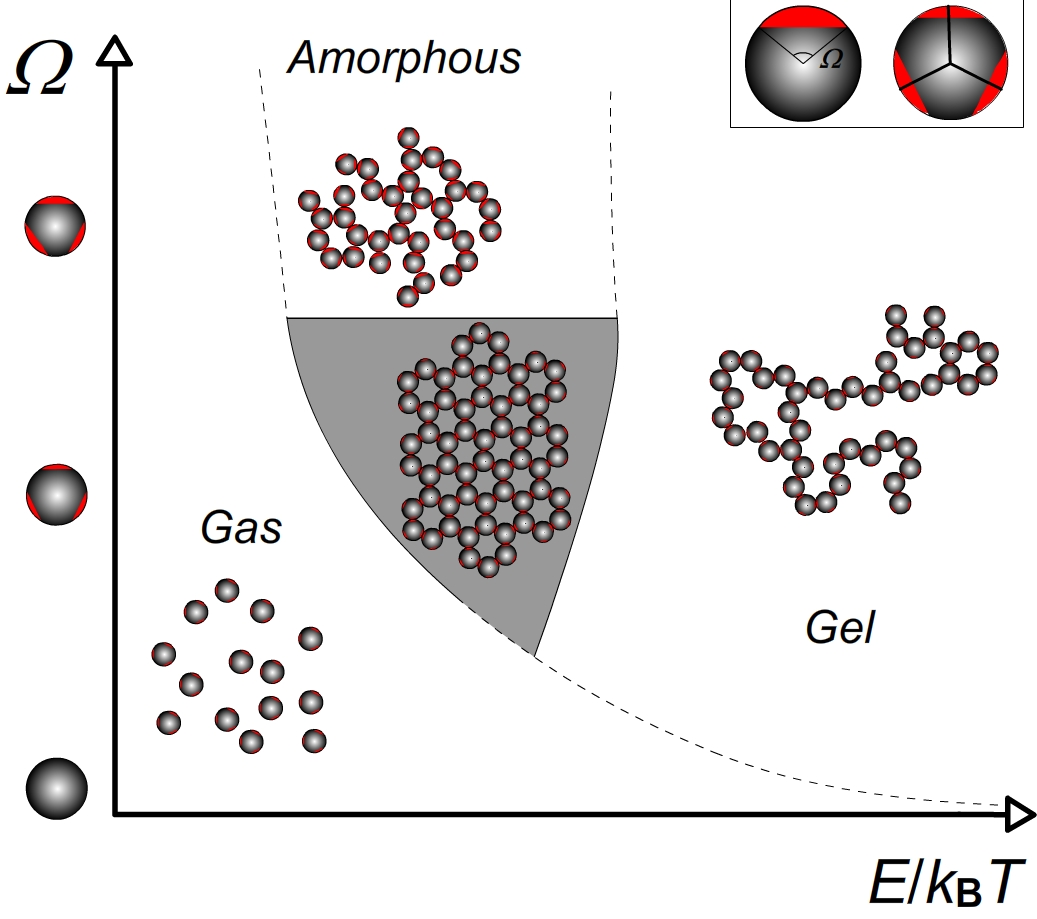}
	\caption{Illustration of a generic self-assembly diagram for patchy spherical particles
	expected to aggregate into a honeycomb lattice. $E$ is the interparticle attraction strength and
	$\Omega$ is the angular size of the patches.}\label{patches}
\end{figure}

Figure~\ref{patches} illustrates the problem for a simple model: spherical particles with attractive patches oriented to form a two dimensional honeycomb lattice.
When the angular size $\Omega$ of each patch is too large, the interaction is not specific enough to select the desired crystal, resulting in amorphous structures originating from the competition of multiple fitting geometries.
If $\Omega$ is too small (too specific), the probability that two particles in close proximity are properly aligned to interact becomes negligible, and the system is found in a gas phase unless a very large interparticle energy $E$ is provided, which in turn drives the system into a gel  phase.
Analogous arguments can be made for the overall strength of the interaction that, for a reasonable patch size, should be neither too strong nor too weak.
The net result is that self-assembly is a very elusive process that requires a careful design and fine tuning of the interparticle interactions, and typically the target region in the interaction space in quite narrow.
 
Our interest is in the problem of ``reverse self-assembly,'' in analogy to the problem of ``reverse protein folding'' in which protein sequences are designed to yield a desired ground-state folded structure~\cite{Yue, Deutsch}.
The problem can be formulated as follows: given an arbitrary final structure, is there an efficient method by which interparticle interactions may be designed so that the particles spontaneously self-assemble into the given structure?

In this paper we use simple physical arguments to develop a numerical procedure capable of sampling the space of interactions, in terms of both patch geometry and binding energy, to generate nanoparticle interactions leading to self-assembly of desired crystal structures.
Without loss of generality, we limit our discussion to spherical particles interacting via anisotropic short-range attractive pair potentials mimicking the hydrophobic interactions driving self-assembly of Janus particles~\cite{Hong2, cacciuto}.    
 
Classical nucleation theory provides a simple framework within which to think about crystal formation.
A free energy gain  $(\mu_c-\mu_f) n$ is associated with the formation of a nucleus of $n$ particles of the crystal phase  at chemical potential $\mu_c$ out of a fluid phase at a larger chemical potential $\mu_f$.
A free energy cost $\gamma A$ is associated to the formation of an interface of  area $A$ ({$\propto n^{2/3}$}) and surface tension $\gamma$ between the two phases.
Minimization the total free energy $\Delta G$ with respect to $n$ leads, for a spherical nucleus,  to a critical size $n_c=\frac{32\pi}{3\rho} \left(\frac{\gamma}{|\Delta\mu|}\right)^3$ ($\rho$ is the equilibrium crystal density). 
Crystal nuclei larger than $n_c$ will grow until the phase transformation is completed, all   others will shrink and vanish.
We argue that a successful strategy for crystal design should take into account the physical properties of the parent fluid phase, and our working hypothesis is that the free energy of crystal nucleation can be exploited to design interactions to target arbitrary crystal structures.  
The main idea is to force a crystal nucleus of desired symmetry to be  in contact with its own fluid and use a numerical procedure to select for those interactions between the particles that minimize the free energy cost required to hold that nucleus in place.
Our scheme consists of two parts: 1) we determine  the optimal shape of the hydrophobic regions ($\Omega$) satisfying the condition stated above, and 2) given $\Omega$, we find  the interaction strength ($E$) for which the system is likely to nucleate into the target (defect-free) crystal phase.  

\section*{SAMPLING THE INTERACTION SHAPE $\Omega$}

Consider a system of $N$ identical particles with a given interparticle potential $U(\Omega_i,r)$ set in a volume $V$.
Define an order parameter $q$ capable of detecting the symmetry of the desired crystal phase.
Grow from the fluid and equilibrate a crystalline nucleus of size $n_0$ using a standard bias Monte Carlo method targeting the size of the largest crystalline cluster in the system, $n$, via a potential  
$V_B(n)=\frac{\kappa}{2} (n-n_0)^2$~\cite{tenwolde2}. 
Set the binding energy among the particles to a sufficiently small value to ensure that the nucleus melts once the bias is removed, and compute from a full simulation in the presence of the bias the average crystal size $\bar n(\Omega_i)$.

Now define a design potential $V_D[\bar n(\Omega_i)]=-\alpha \bar n(\Omega_i)$, where $\alpha$ is a numerical constant. 
At this point the idea is to sample over the space of interactions using  $V_D$ as a driving force.
Specifically, we generate an alternative (trial) shape for the interaction between any two particles in the system $\Omega_j=\Omega_i+\Delta\Omega$ and repeat the previous steps to obtain a new estimate for $V_D[\bar n(\Omega_j)]$.
$\Omega_j$ is accepted or rejected based on a standard Metropolis criterion, thus ensuring that the $\Omega$ will be driven towards values that maximize the size $\bar n$, i.e. minimize the load requested of the bias to hold the crystal in place.
 
\section*{SAMPLING THE INTERACTION STRENGTH $E$}

Unfortunately, our method does not allow easy measurement of the height of the nucleation barrier given an interaction strength $E$.
This is mostly because the surface tension between the crystal and the fluid phase is unknown.
Nevertheless, we have direct access to the slope of the free energy barrier.
Therefore, although one cannot design the system to comply with a specific nucleation rate $\nu$, by modulating $E$ one can design the size of the critical nucleus $n_c$. 
A critical nucleus that is too small will result in the almost instantaneous nucleation of several crystallites that will form defects and grain boundaries as they meet while growing.
The opposite scenario will lead to absence of crystallization within the experimental time frame.
For the systems we have examined, we find that $n_c \sim 15-30$ results in nucleation events that are quick, yet sufficiently rare to prevent formation of multiple crystals.
The choice of $n_c$ may require a few iterations depending on the details and the size of the system.

The strategy behind the design of the critical nucleus size is analogous to that described in the previous case, except that the design potential in this case has a harmonic functional form defined as $V_D[\bar n(E_i)]=\alpha (\bar n(E_i)-n_c)^2$, and we sample over the interaction strength $E$.
Minimizing $V_D$ implies that the system will be driven towards that value of $E$ ($E_c$) for which the nucleus has the same probability of growing or shrinking.
This condition guarantees that the system is at the top of the nucleation free energy barrier, and that $n_c$ has indeed become the critical nucleus by definition.

{Note that in principle, the interaction geometry could have an arbitrary number of parameters that could all be optimized simultaneously; however, it is crucial that the optimization of the geometry precedes the optimization of the strength of the potential.
In fact, it is mandatory for the nucleus to be precritical in order for the geometry optimization scheme to be effective.}

\section*{NUMERICAL TESTS}

As a proof of concept for our method we consider the design of two distinct crystal structures for which we can guess the solution in the interaction space and   know how to define an order parameter $q$: a simple cubic crystal (SC) and a two dimensional sheet with square symmetry  embedded in a three dimensional environment (2SQ).

For both systems we adopt the Kern-Frenkel model~\cite{Kern}. 
Particles are described as hard spheres of diameter $\sigma$ interacting with a short-range attractive interaction that is turned on whenever hydrophobic regions (the patches) on different particles face each other.
For each  pair of particles $i$ and $j$ with patch indices $\alpha$ and $\beta$, the interaction is defined as 
	\begin{equation}
		\label{eq1}
		u(\mathbf{r}_{ij})=u_{\rm SW}(r_{ij}) \sum_{\alpha,\beta}f^{\alpha\beta}(\Theta_{ij})
	\end{equation}
where $ u_{\rm SW}(r_{ij})$ is a standard attractive isotropic square well potential of depth $\varepsilon$ and range $1.15\sigma$, and $f^{\alpha\beta}(\Theta_{ij})$ depends on the particles' mutual orientations and is defined as
	\begin{equation}
		\label{eq2}
		f^{\alpha\beta}(\Theta_{ij})= \left\{
		\begin{array}{ll}
			1 &\mbox{if \, }\left\{
			\begin{array}{lll} \mbox{ \,\,\,\,\,\,\, \,\,\,$\hat{\bf r}_{ij}\cdot\hat{\bf e}_{\alpha}>\cos \theta$ } & \begin{array}{l}\end{array} &\\ 
    			\mbox{ and $\hat{\bf r}_{ji}\cdot\hat{\bf e}_{\beta}>\cos\theta$ } & \begin{array}{l} \end{array}
            \end{array} \right.\\
		0 & \mbox{otherwise}
		\end{array} \right.
	\end{equation}

Here $\theta$ is the angular size of the hydrophobic regions (selected to be all identical in size and circular in shape), $\hat {\bf r_{ij}}$ is the unit vector along the direction of the interparticle separation, and $\hat{\bf e}_{\alpha}$ is the unit vector connecting the center of a particle to the center of the patch $\alpha$ on its surface.
 
In these simple systems $\theta$ and $\varepsilon$ are the design parameter we need to tune for self-assembly to take place.
All of our simulations are performed in the $NVT$ ensemble using a minimum of 256 particles in a box with periodic boundary conditions.
A good order parameter to detect SC crystals is the standard local bond order based on spherical harmonics, $\bar{q}_4$~\cite{ronchetti,tenwolde2,auer}. 
Given a particle  $i$, we compute
\begin{equation}
Q_{4m}(i)=  \frac{1}{N_b(i)}    \sum_{j=1}^{N_b(i)} Y_{4m}({\bf r}_{ij})\,
\end{equation}
where $j$ runs over the $N_b(i)$ neighbors of  particle $i$, from which a rotationally invariant order parameter correlating the orientation 
of neighboring particles $i$ and $j$ can be defined as
\begin{equation}
{\bf q}_4(i)\cdot {\bf q}_4(j)= \sum_{m=-4}^4 Q_{4m}(i) \cdot Q^*_{4m}(j) 
\end{equation}
Once averaged over all neighbors $j$, the resulting quantity $\bar{q}_4$ is compared to a cutoff, $q_{\rm cut}$,
to decide whether a particle can be tagged as crystalline or not.
 
For the 2SQ case we used $\bar{q}_4$ with the added constraint that a particle must have interactions with no more than four neighbors in order to be considered ``crystalline.'' The location of the patches automatically prevents the formation of SC crystals in this case. The insets in Figure 2 sketch 
the patch positions over the particles.

\begin{figure*}
	\subfigure[]{
		\includegraphics[width=0.45\textwidth]{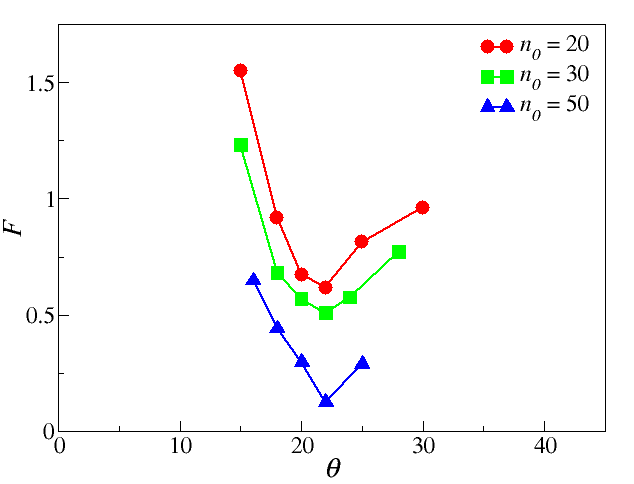}
		\put(-68,25){\includegraphics[width=63px]{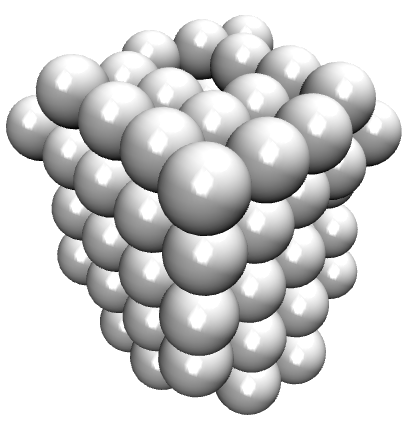}}
		\put(-185,25){\includegraphics[width=45px]{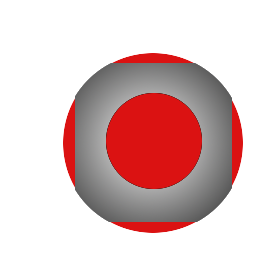}}
		\label{force:SC}
	}
	\subfigure[]{
		\includegraphics[width=0.45\textwidth]{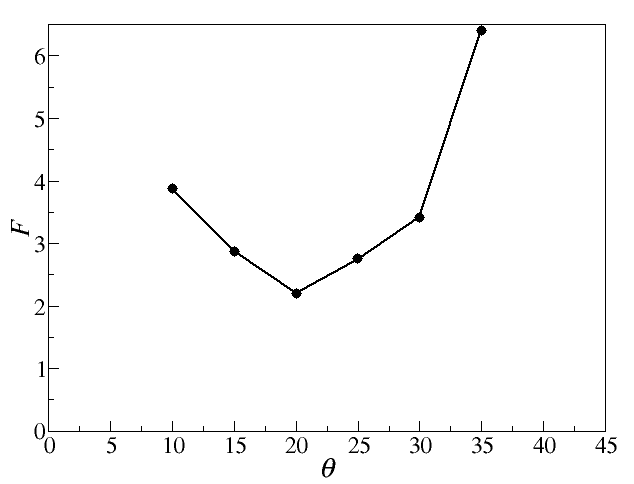}
		\put(-80,25){\includegraphics[width=75px]{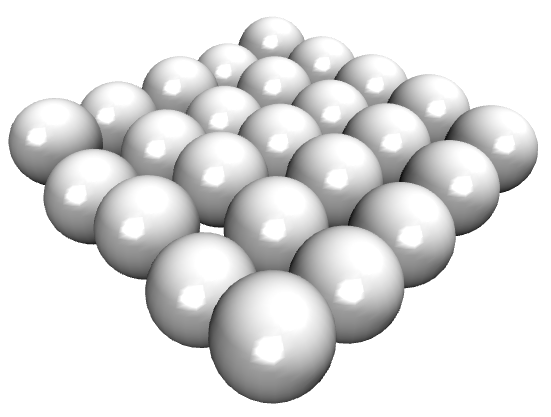}}
		\put(-190,25){\includegraphics[width=45px]{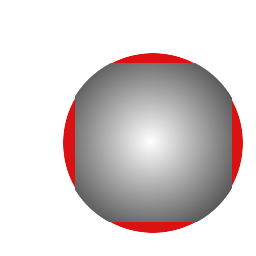}}
		\label{force:2SQ}
	}
	\caption{Force vs. $\theta$ for \subref{force:SC} SC and \subref{force:2SQ} 2SQ crystals.  The insets show snapshots of the target crystals, and  sketches of the locations of the patches in our particle model. In
\subref{force:SC}, the different lines represent data obtained by imposing different nucleus sizes $n_0$ as indicated
in the legend.}
\end{figure*}

Figures~\ref{force:SC} and ~\ref{force:2SQ} illustrate how the force $F(\theta)=-\kappa (n-n_0)$ required to hold a nucleus of $n_0$ particles immersed in its fluid phase depends on the size of the circular regions $\theta$ for the SC  and the 2SQ crystals respectively, and specifically Fig.~\ref{force:SC} also shows that the optimal value is fairly independent of  the particular size of the nucleus  $n_0$  held in contact with the fluid.

\begin{figure*}
	\subfigure[]{
		\includegraphics[width=0.475\textwidth]{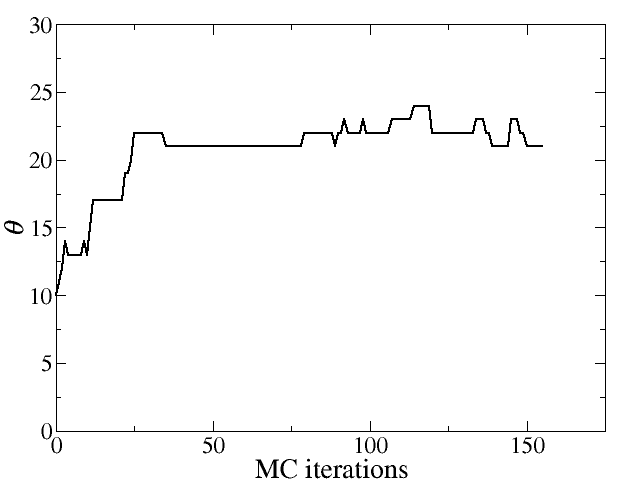}
	}
	\subfigure[]{
		\includegraphics[width=0.475\textwidth]{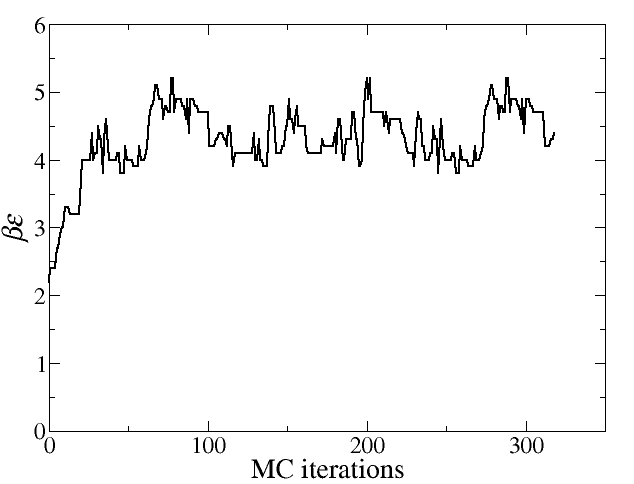}
	
	} 
	\caption{Monte Carlo trajectories in the space of interactions for the design of the simple cubic crystal. In (a) the shape of the patches defined by the solid angle $\theta$ is allowed to fluctuate while keeping the strength of the interaction $\varepsilon$ constant. In (b) $\varepsilon$ fluctuates while keeping $\theta$ constant and at the optimal value found in (a).}	\label{trajectory} 
\end{figure*}

The corresponding simulations were performed at densities of $\rho_{SC} = 0.1$ and $\rho_{2SQ} = 0.01$, binding strength $\varepsilon_{SC}=3.5k_{\rm B}T$ and $\varepsilon_{2SQ} = 5.75k_{\rm B}T$, and a harmonic bias potential of spring constant {$k_{SC} \simeq 0.2k_{\rm B}T$} and {$k_{2SQ} = 0.4 k_{\rm B}T$}.
Clearly, $F(\theta)$ is a sufficiently sensitive parameter to discriminate among the different angular sizes,  and presents in both cases a distinct optimal value; $\theta^*_{SC}\simeq 22^{\circ}$ and $\theta^*_{2SQ}\simeq 20^{\circ}$.
Figure~\ref{force:SC} also shows that the optimal value is fairly independent of  the particular average size 
$n_0$ of the nucleus held in contact with the fluid.
Figure~\ref{trajectory} shows how the location of $\theta^*$ and $\varepsilon^*$ can be obtained automatically by using the Monte Carlo scheme in the space of interactions.

It should be stressed that the minimization of $V_D[\bar n(\Omega_i)]$ can be achieved using any minimization algorithm; nevertheless, we find that the Monte Carlo scheme allows us to use shorter simulations, for each trial $\theta_i$, than what would be necessary for other direct minimization schemes.
The reason is related to the precision of the estimate of $\bar n$ for relatively short trajectories that could be 
over- or underestimated.
This could lead to fictitious local minima, which could trap a direct minimization scheme, but are easily overcome with a standard Monte Carlo method.

\begin{figure*}
	\subfigure[]{
		\includegraphics[width=0.45\textwidth]{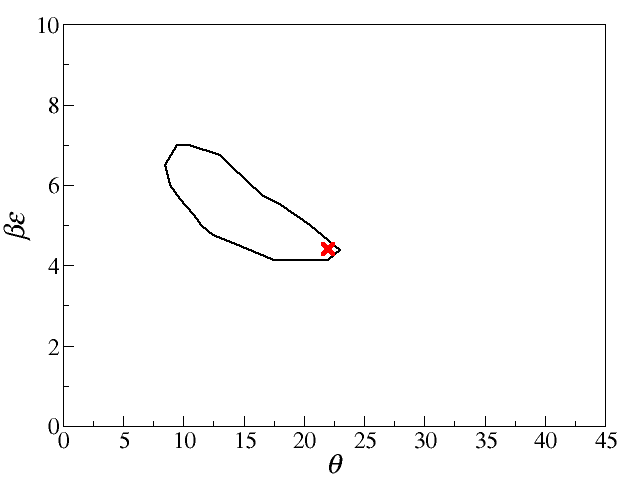}
		\label{phase:SC}
	}
	\subfigure[]{
		\includegraphics[width=0.45\textwidth]{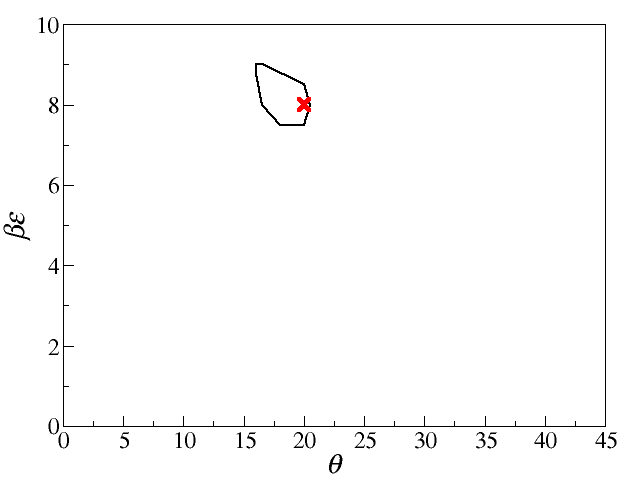}
		\label{phase:2SQ}
	}
	\caption{Phase diagrams for \subref{phase:SC} SC and \subref{phase:2SQ} 2SQ crystals.  Lines show the border around the phase region in which particle form the desired crystal; the points marked by an `X' are the $(\theta, \beta\varepsilon)$ combinations found to be optimal by our method.}\label{phase}
\end{figure*}

In order to check our method, $\varepsilon-\theta$ phase diagrams were constructed using the traditional, ``forward'' method of trial-and-error{, running Monte Carlo simulations for $10^7$ steps and determining whether crystallization occured}.  As shown in Figure~\ref{phase}, the parameters detected by our methods are within the crystallization regions for the two target crystals.
Unsurprisingly, the result falls roughly to the high-$\theta$, low-$\varepsilon$ edge of the crystallization region; recall that $\theta$ was selected using a value of $\varepsilon$ too low for crystallization, which would be expected to result in a larger $\theta$ (note the roughly inverse $\varepsilon-\theta$ relationship in Figure~\ref{phase}); $\varepsilon$ was then selected using the $\theta$ found in the first step.

\section*{LIMITATIONS}
It is important to discuss the limitations of the method.
First of all, in its actual formulation, our method only works for systems  that will self-assemble
into an infinitely large aggregate via the process of nucleation. It is not obvious how to generalize it
to include self-assembly into finite size aggregates such as for instance viral capsids.

The second limitation is that although the method provides a solution to the reverse self-assembly problem,
there is no guarantee that the solution is the optimal one. This is because our method forces the 
nucleation process to follow the classical route, i.e. the forming nucleus has the same structure as the 
target crystal; however, there are several   examples~\cite{cacciuto,tenwolde,cacciuto22,whitelam,kumar}  where the nucleation barrier may be 
lowered by following a more complex dynamical pathway that may include metastable states having 
different symmetry than that of the target crystal. For instance, it is possible to imagine that 
the formation of the SC crystal could benefit from an additional weak, non-specific, isotropic interaction, on top 
of that provided by the patches, that may initially lead the system into a high-density metastable fluid
phase from which nucleation into the final structure may proceed at a faster rate than that predicted otherwise.

\begin{figure}
\includegraphics[width=.45\textwidth]{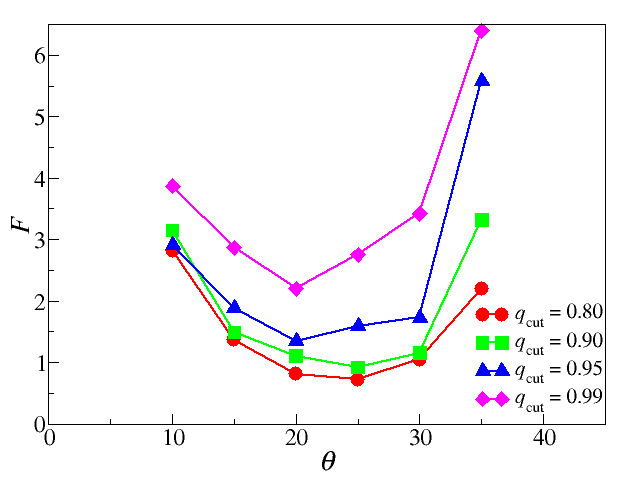}
\caption{$F$ vs. $\theta$ diagrams.  The dependence of the method on the order parameter used.  Different curves correspond to different values of the  cutoff $q_{\rm cut}$.} \label{orderparameter}	 
\end{figure}

Finally, it is crucial to develop a good order parameter $q$ to describe the desired crystal structure.   
Figure~\ref{orderparameter} illustrates how an inefficient order parameter may lead to fictitious minimization in the space of interactions
while designing the 2SQ crystal.
The different lines in the $F$ vs. $\theta$ diagram represent different values of $q_{\rm cut}$ (defining how restrictive the order parameter is) from $0.8$ to $0.99$.
We find that a cutoff in the order parameter of at least 0.97 is required to adequately distinguish between the square and hexagonal symmetries for large values of $\theta$. The curves related to the less restrictive order parameters would in fact misleadingly indicate a flatter
bottom of the curve, while in reality we find that any angle larger than $\sim25^{\circ}$ will lead to nucleation into a two-dimensional crystal with hexagonal symmetry. 
Adding an energy penalty to  prevent arrangements compatible with the competing six-fold symmetry (apart from 
imposing a limit to the number of neighbors) may also be considered as a means of  improving the design procedure.

\section*{BEYOND THE KERN-FRENKEL MODEL}
Here we propose a more general model to describe interparticle interactions that we name the {\it Adaptive Pixel Model}.
The idea stems from the need to devise a  way of sampling over different geometric patterns (beyond circular) 
in search of those which can efficiently hold the single components into a desired target structure.
The first step is the discretization of the surface of the particle.

\begin{figure}
		\includegraphics[width=0.5\textwidth]{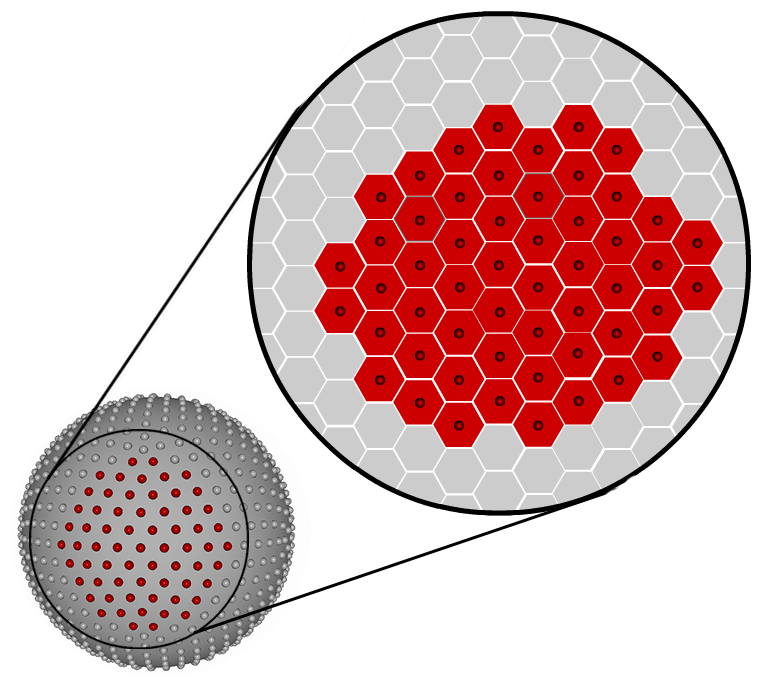}
		\caption{  Illustration of the structure of the {\it Adaptive Pixel  Model}. {\it on} pixels are depicted in red while {\it off} pixels are in gray. The magnification in the top image shows the Voronoi tessellation around the pixels (computed as described in the text). The effective geometry of the active sites in this representation is a hexagon. }
		\label{pixels}
	\end{figure}

For spherical building blocks, we cover the surface of each particle with a large number, $N_p$, of regularly-spaced 
interaction sites (pixels) as illustrated in Fig.~\ref{pixels}.
A good arrangement of the pixels can be obtained by using the spherical triangulation provided by an $(n,m)$ delta-icosahedron~\cite{geom},
and $N_p$ is selected depending on the complexity of the target structure.
Euler's theorem~\cite{geom} imposes the following constraint on  $N_p$: $N_p=10(n^2+nm+m^2)+2$~\cite{geom}, 
where $n$ and $m$  indicate that one has to move $n$ pixels  along the row of neighboring bonds on the sphere, and then after a
turn of $120^{\rm o}$, move for $m$ extra pixels. 
 
In the simplest version of the model, to each pixel $k$ on a particle $i$, is assigned a variable $s_{ik}$ which has a binary character, $s_{ik}\in\{1,0\}$ depending on whether that  interaction site is switched {\it on} or {\it off}.
Whenever two particles $i$ and $j$ are within a given distance of each other, the axis between them, $r_{ij}$, is calculated.
If the nearest digit to the point where $r_{ij}$ crosses the surface of each particle is {\it on}, then the particles feel an overall  short-range attractive interaction.
Pixels on the same particle do not interact with each other. The interaction pair potential between any two particles, $i$ and $j$, of diameter $\sigma$, set at a distance $r_{ij}$ from each other, then takes the form
	\begin{equation}
		V(r_{ij}) = V_{\rm HS} + \begin{cases} 	-s_{ik}s_{jl}\varepsilon		&	\textrm{if } |r_{ij}| \leq r_0 \\
											0 							&	\textrm{otherwise}\end{cases}
		\label{interaction}
	\end{equation}
where $s_{ik}$ and $s_{jl}$ are the binary variables corresponding to the digits intersected by $r_{ij}$ on particles $i$ and $j$, respectively, as described above.
Excluded volume between the particles is enforced via a standard hard-sphere potential, $V_{\rm HS}$. 

The main advantage of this setup is that once particles are held into place at given positions, the geometry of the interacting regions emerges as a result of a simple Monte Carlo simulation on $s_{ik}$ which samples different states according to Eq.~\ref{interaction}.
Crucial to the efficiency of the model, is the independence of the interaction strength of the total area of the attracting region.
This condition, also assumed in the Kern-Frenkel model, is appropriate when considering interactions
that have a range of action that is small compared to the colloidal diameter (($r_0-\sigma)\lesssim 0.15\sigma$), and allows us to circumvent the overwhelming cost related to the computation of  the $N_k^2$ distances between the pixels.
Furthermore, as the relative distances of on-particle pixels are frozen and only active pixels need to be tracked, it is possible to perform the
search of the nearest pixel to any point on the sphere very efficiently. This is achieved by creating a cell list 
over the spherical particle surface in $\theta$ and $\phi$ (the spherical coordinates), and by associating to each cell the identity of the
nearest pixel. This is equivalent to generating a discrete Voronoi tessellation~\cite{geom} of the spherical surface based on the pixel locations (see sketch in Figure~\ref{pixels}), which needs to be performed only once at the beginning of the simulation.
Any shape for the interaction regions can be achieved by simply switching {\it on} or {\it off} pixels or groups of pixels.

\section*{CONCLUSIONS}

In this paper we have proposed a simple two-step method for the problem of reverse self-assembly.
The idea is to exploit the curvature of the nucleation free-energy barrier to sample and select optimal 
inter-particle interactions for self-assembly into a target structure. We presented numerical simulations 
to test the efficacy of our method, and discussed in detail its limitations and its potential. 
{These simulations show that our method reduces the time to solve the problem of determining optimal interaction parameter from on the order of weeks (for trial-and-error approaches) to hours.}

Finally, we proposed a new model, the {\it {Adaptive Pixel Model}}, by which almost any interaction geometry can be realized in a simple and efficient way.
It should be stressed that our method is not limited to spherical particles but can be applied to any particle shape. 
In principle, particle shape could be introduced as a new parameter in the interaction space and be sampled over using the scheme proposed in
this paper.

Although our method does not capture the dynamical subtleties of the crystal formation process, it does provide 
a very efficient way of screening over a large number of given interaction geometries that can be mapped onto the pixels.
Efficient ways of sampling the interaction space could be obtained using genetic algorithms that can be used to evolve  optimal  interaction patterns given a set of initial shapes. Work in this direction is currently under investigation and will be published elsewhere.

\section*{ACKNOWLEDGMENTS}
This work was supported by the National Science Foundation
under CAREER Grant No. DMR-0846426.

\end{document}